# A snapshot of electrified nanodroplets

Steve Arscott[1,a)], Cédric Descatoire[1], Lionel Buchaillot,[1] and Alison E. Ashcroft[2]

[1]*Institut d'Electronique, de Microélectronique et de Nanotechnologie (IEMN), CNRS, Université de Lille, Avenue Poincaré, Cite Scientifique, Villeneuve d'Ascq 59652, France.* [2]*Astbury Centre for Structural Molecular Biology, University of Leeds, Leeds LS2 9JT, United Kingdom.*

We investigate the size distribution of electrically charged nanodroplets. The droplets were generated using nano- and micro- scale silicon tips. A brief voltage pulse results in a "snapshot" of charged nanodroplets on a metal surface. Atomic force microscopy (AFM) of the snapshot revealed that certain droplet diameters are favored suggesting droplet fission due to Rayleigh instability at nanometer length scales. The most occurring droplet diameters are 85.9(4.1) nm and 167.1 nm (9.7 nm) and for nano- and micro- scale tips respectively indicating that the tip size determines deposition resolution.

a) steve.arscott@iemn.univ-lille1.fr

Electrically charged droplets have an important place in physical sciences[1], life sciences[2], technology[3], agriculture[4] and natural phenomena[5]. In 1882 Lord Rayleigh[6] predicted that such droplets will be unstable if the electrical charge they carry exceeds a value given by $\sqrt{8\pi^2 \gamma \varepsilon_0 d^3}$, where $d$ is the droplet diameter, $\gamma$ is the surface tension and $\varepsilon_0$ is the permittivity of free space. Coulomb fission, due to this instability, has recently been observed for micrometer-sized droplets[7]. We demonstrate here an original technique which can produce a "snapshot" of charged nanometer-sized droplets under Rayleigh instability. A nanometer-scale silicon tip is used to deposit[8] nanodroplets onto a metal screen. Atomic force microscopy of the snapshot suggests Coulomb fission occurring at these length scales; the results also indicate that tip size determines the deposition resolution.

The experimental set-up is shown in Fig. 1(a). The tips are composed of two triangular cantilevers, attached to a silicon support chip, which define a micrometer capillary slot leading up to a nanometer scale capillary slot at the apex of the tip[9] [Fig. 1(b)]. The tip apex has a nanometer scale channel which was defined using focused ion beam milling. Two nanometer-scale tips were tested having channel widths of 350 nm (nanoA) and 400 nm (nanoB) and as a comparison, micrometer sized tips[10] were also tested (microA and microB); these had channel width of 1 µm [Fig. 1(c)]. The tip is brought into proximity to a chromium (thickness = 200 nm) coated silicon wafer using an *xyz* positioning stage; the tip-to-plane distance was set to ~100 µm. The tip is loaded with the liquid ($v$ ~ 5 µL) composed of a 75:25 *v/v* deionized water ($\rho$>10 MΩ cm)-methanol mixture with nitric acid ($c$ = 0.001 M); spontaneous capillary filling[11] of the capillary slot ensures that liquid fills to the apex of the tip. A gold wire (diameter = 250 µm) is inserted into the droplet to serve as an electrode. A voltage pulse (0-200 V for 10 ms) is applied to the liquid which can cause electrospraying at the apex of the tip[12]. For a point-to-plane electric field distribution[13] the field decreases rapidly in a non-linear fashion in the axial direction away from the point. During the experiments the electric field at the tip was large enough to produce electrospraying[9,12] but decreases rapidly to below the value of break-down field of air; arcing behaviour[14] was not observed in the current-voltage sweeps but rather a characteristic electrospray current-voltage[15]. The emitted charged droplets are attracted towards the chromium coated silicon wafer which is grounded [Fig. 1(a)]. The time-of-flight of the droplets is calculated to be of the order of tens of microseconds by numerically resolving a point-to-plane electric field model[13]. Break-up of a cone-jet[16,17] results in the formation of the electrospray plume as the charged droplets repel each other. During the time-of-flight, the charged droplets which exceed the Rayleigh criterion[6] will undergo Coulomb fission which has been the subject of much research[8]. Characterizing the distribution of micrometer-sized charged droplets can be done using optical means[7,8,18,19]. In contrast, the method demonstrated here can characterize *individual droplets* according to their size and occurrence which could lead to a deeper insight into the spray. Literature is scarce concerning imaging individual charged nanodroplets originating from an electrospray; scanning electron microscopy and near-field microscopy techniques have been used to characterize nanodroplets deposited onto surfaces[20-24] and we use near-field microscopy techniques here to characterize the snapshot of charged nanodroplets.

Fig. 2(a) shows near-field microscopy image obtained by tapping-mode atomic force microscopy (AFM) imaging (Bioscope, Veeco USA) of the chromium surface following experiments with the nano-scale tip "nanoA". The traces are typically composed of a central area ~10 µm in diameter surrounded by a halo[13] of much smaller sub-micrometer diameter satellite "spots"[13] characteristic of electrostatic repulsion[19]. Each satellite spot [Fig 2(b)] has a central "bump" characteristic of a non-volatile residue[20] possibly due to impurities (rainbow colors in Fig. 2(b), surrounded by a flat "plateau" region [brown in Fig. 2(b)] not reported in previous studies[21-25]. The overall spot profile is not that of a spherical cap; this in contrast to other work[21,23,25] who measured spherical cap profile nanodroplets by evaporation/AFM techniques[21,23] and dip-pen/AFM methods[25]. The trace diameters are ~50 µm (Inset of Fig. 2(c) which shows that smaller droplets are found towards the periphery of the trace), implying an electrospray plume angle of around 30°; this is consistent with previous observations[13,19].

By analyzing these images we can produce Fig. 2(c) which shows a plot of the normalized spot number ($n/N_t$) versus spot diameter $w$ for four tips. It is apparent from Fig. 2(c) that certain spot diameters are favored and occur in bunches (see steps in Fig. 2(d) indicated by black arrows). By statistically binning the experimental data presented in Fig. 2(c) using a suitable class interval we can produce histograms for the spot width $w$ versus droplet occurrence $N$; the result of this is shown in Figs. 2(e) and 2(f). To explain these results let us recall that a charged droplet distribution has few large droplets (the fissility[7] $X = q^2/8\pi^2\varepsilon_0\gamma d^3$ is large) and few small droplets ($X$ is small) but is composed of many droplets of distinct radii in the vicinity of $X$~1; the distributions in Figs. 2(e) and 2(f) are indicative of this. Traces produced using the nano-scale tips result in a minimum spot width variation of 77.9-1101 nm (nanoA) and 99.8-631 nm (nanoB) whereas micro-scale tips result in 157-2578 nm (microA) and 172-2490 nm (microB).

In order to interpret our observations Fig. 3 illustrates a charged droplet impinging on the metal surface where local oxidation[26] and desposition[27] can occur; the result of this is a modification of the surface resulting in a spot having a diameter $w$. As a first approximation we can relate $w$ to the original impinging droplet diameter $d$ by using a model based on the macroscopic wetting contact angle $\theta$ of the water-methanol-nitric acid mixture on a Cr surface; although it should be noted that a more accurate model would require nanodroplet wetting[21] and charge effects to be taken into account. However, in a first approximation as $d = \alpha w$ where $\alpha^3 = (2 - 3\cos\theta + \cos^3\theta)/4\sin^3\theta$ and $\theta$ was measured to be 24.7 (±2)° using a contact angle meter (Kruss, Germany) we determine the parameter $\alpha$ to be 0.437. This enables a calculation of the experimental droplet diameters ($d_{\exp}$) shown in Tables I-IV. The experimental droplet diameters $d_{\exp}$ are mean values calculated over the class interval. The observations can be analyzed in terms of (i) droplet size distribution and (ii) volume flow rate.

Firstly, in terms of the droplet size distribution although electrically charged droplet distribution populations are known to be highly complex[28] a simple fission model based on droplet volume halving can be compared with the experimental results. This simplistic model can check for a signature of droplet break-up at these length scales, presumably due to Rayleigh instability[6]. By choosing an highly occurring droplet diameter $d_n$ (red and blue underlined bold values in the Tables I-IV) observed in the experimental AFM data we can calculate a set of droplet diameters based on simple volume halving: $V_{n+1} = V_n/2 \Rightarrow d_{n+1} = d_n/\sqrt[3]{2}$; these calculated droplet diameters $d_{\mathrm{cal}}$ are shown in Tables I-IV. There are several experimental droplet diameters which correspond very well to the simple volume halving fission model; one can presume that this indicates a signature of Coulomb fission and gives evidence for nanodroplet splitting due to Rayleigh instability at these length scales although a full understanding of the droplet distribution population requires more complex models[28]. The nano-scale tips produce smaller droplets than the micro-scale tips. Interestingly there are some common diameters: ~53 nm (for the nano-scale tips) and ~166 nm and ~202 nm (for the micro-scale tips). Also, by making the assumption that the most occurring droplet diameters have $X=1$ we can calculate the number of unit charges to be 902 droplet$^{-1}$ (6.22×10$^{-3}$ C m$^{-2}$) and 2446 droplet$^{-1}$ (4.46×10$^{-3}$ C m$^{-2}$) for the nano-scale and micro-scale tips respectively.

Secondly, a volume flow rate $Q$ can be determined from the measurements. The measurement sector angle $\psi$ for the nano-scale tip (nanoA) was equal to 31° (corresponding to 498 spots in this AFM image or ~5700 spots in the total trace) whereas $\psi$ for the micro-scale tip (microA) was 25° (corresponding to 276 spots in this AFM image or ~3900 spots in the total trace). $Q$ can be calculated by summing all droplet volumes in the AFM trace and multiplying by a factor which takes into account the measurement sector $\psi$ of the total trace. In this way, we determine $Q$ to be equal to 0.86(±0.1) nL min$^{-1}$ (nanoA) and 3.72(±0.3) nL min$^{-1}$ (microA). By using the physical properties of the water-methanol-nitric acid mixture: surface tension $\gamma$ (~47 mJ m$^{-2}$)[29], density $\rho$ (~844 kg m$^{-3}$)[30] and conductivity $\sigma$ (~1.26 µS m$^{-1}$); we can compare our findings here to validated models and experiments in the literature[8]. The current $I$ was measured to be 4.5 nA (nanoA) and 9.6 nA (microA). The scaling law[31] $Q \sim I^2/\gamma\sigma$, which has been rigorously verified[8], yields a constant of proportionality equal to 4.21×10$^{-5}$ and 3.98×10$^{-5}$ for the nano- and micro-scale tips. For varicose "Rayleigh" break-up[32,33], the most likely occurring here, the droplet diameter ($d = 2.27\pi^{-2/3}Q^{1/2}(\rho\varepsilon_0/\gamma\sigma)^{1/6}$) can be calculated[32] to be 89.6(±5.1) nm and 186.4(±7.4) nm for the nano and micro tips respectively; these values are comparable with the experimental most occurring values in Tables I and III.

In conclusion, we demonstrate a method which can produce and characterize a snapshot of nanodroplets using nanofabricated silicon tips and AFM. The study provides evidence for Coulomb fission of nanodroplets. This seems the most likely explanation for our observations as the experimental results are not characteristic of arcing[14], electrowetting[34], droplet impact[35] or evaporation[36]. The approach could be useful for testing existing models[8,28], characterizing new phenomena such as catastrophic droplet breakup[37]. Also, the observations have implications for nanotechnology[38] as the tip size is seen to determine the deposition resolution.

(1) Millikan, R.A. *Phys. Rev.* **1913**, *2*, 109–143.
(2) Fenn, J. B.; Mann, M.; Meng, C. K.; Wong, S. F.; Whitehouse, C.M. *Science* **1989**, *246*, 64–71.
(3) Hines, R. L. *J. Appl. Phys.* **1966**, *37*, 2730–2736.
(4) Law, S. E. *J. Electrostatics* **2001**, *51-52*, 25-42.
(5) The electrical nature of storms, D.R. Macgorman & W. D. Rust, Oxford University Press USA.
(6) Rayleigh, Lord *Phil. Mag.* **1882**, *14*, 184–186.
(7) Duft, D.; Achtzehn, T.; Müller, R.; Huber, B. A.; Leisner T. *Nature* **2003**, *421*, 128.
(8) Ganan-Calvo, A. M.; Montanero, A. M. *Phys. Rev.* E **2009**, *79*, 066305-18.

(9) Arscott, S.; Troadec, D. *Nanotechnology* **2005**, *16*, 2295-2302.

(10) Arscott, S.; Le Gac, S.; Rolando, C. *Sens. Act.* B **2005**, *106*, 741-749.

(11) Brinkmann, M.; Blossey, R.; Arscott S.; Druon, C.; Tabourier, P.; Le Gac, S.; Rolando, C. *Appl. Phys. Lett.* **2004**, *85*, 2140-2142.

(12) Arscott, S.; Troadec, D. *Appl. Phys. Lett.* **2005**, *87*, 124101-3.

(13) Gañán-Calvo, A. M.; Lasheras, J. C.; Dávila, J.; Barrero, *J. Areosol. Sci.* **1994**, *25*, 1121-1142.

(14) Morrow, R. *J. Phys. D: Appl. Phys.* **1997**, *30*, 3099–3114.

(15) Van Berkel, G. J.; Zhou, F. *Anal. Chem.* **1995**, *67*, 2916-2923.

(16) Taylor, G. *Proc. R. Soc. Ser.* A **1964**, *280*, 383–97.

(17) Gañán-Calvo, A. M. *Phys. Rev. Lett.* **1997**, *79*, 217-220.

(18) Bachalo, W. D.; Houser, M. J. *Optical Engineering* **1984**, *23*, 583-590.

(19) Gomez, A.; Tang, A. *Phys. Fluids* **1994**, *6*, 404-414.

(20) Fittschen, U. E. A.; Bings, N. H.; Hauschild, S.; Forster, S.; Kiera, A. F.; Karavani, E.; Fromsdorf, A.; Thiele, J. *Anal. Chem.* **2008**, *80*, 1967-1977.

(21) Checco, A.; Guenoun, P.; Daillant, J. *Phys Rev. Letts.* **2003**, *91*, 186101-1.

(22) Miller, T. C.; Havrilla, G. J. *X-Ray Spectrom.* **2004**, *33*, 101–106.

(23) Ma, J.; Jing, G.; Chen, S.; Yu, D. *J. Phys. Chem.* C **2009**, *113*, 16169–16173.

(24) Li, D.; Marquez, M.; Xia, Y. *Chem.Phys. Lett.* **2007**, *445*, 271–275.

(25) Jung, Y.C.; Bhushan, B. *J. Vac. Sci. Technol.* A **2008**, *26*, 777-782.

(26) Dagata, J. A.; Schneir, J.; Harary, H. H.; Evans, C. J.; Postek, M.T.; Bennett J. *Appl. Phys. Lett.* **1990**, *56*, 2001-2003.

(27) Wilhelm, O.; Madler, L.; Pratsinis, S. E. *J. Aerosol Sci.* **2003**, *34*, 815–836.

(28) Vazquez, R.; Gañán-Calvo, A. M. *J. Phys. A: Math. Theor.* **2010**, *43*, 185501-22.

(29) Vaquez, G.; Alvarez, E.; Navaza, J. M. *J. Chem. Eng. Data* **1995**, *40*, 611-614.

(30) CRC Handbook of Chemistry and Physics, 69[th] Edition, Edited by R. C. Weast, CRC Press, Florida, USA.

(31) Gañán-Calvo, A. M.; Barrero, A.; Pantano-Rubiño, C. *J. Aerosol Sci.* **1993**, *24*, S19-S20.

(32) Gañán-Calvo, A. M.; Dávila, J.; Barrero, A. *J. Aerosol. Sci.* **1997**, *28*, 249-275.

(33) Hartman, R. P. A.; Borra1, J.-P.; Brunner, D. J.; Marijnissen, J. C. M.; Scarlett, B. *J. Electrostatics* **1999**, *47*, 143-170.

(34) Mugele, F.; Herminghaus, S. *Appl. Phys. Lett.* **2002**, *81*, 2302-2305.

(35) Gamero-Castan, M.; Torrents, A.; Valdevit, L.; Zheng, J-G. *Phys. Rev. Lett.* **2010**, *105*, 145701-4.

(36) Deegan, R. D.; Bakajin, O.; Dupont, T. F.; Huber, G.; Nagel, S. R.; Witten, T. A. *Nature* **1997**, *389*, 827-829.

(37) Raut, J. S.; Akella, S.; Kumar Singh, A.; Naik, V. M. Langmuir **2009**, *25*, 4829-4834.

(38) Jaworek, A.; Sobczyk, A. T. J. Electrostatics **2008**, *66*, 197–219.

FIG. 1. Deposition of nanodroplets onto a metal screen. (a) experimental set-up showing the nano/micro- machined silicon tip, the liquid (water/methanol/nitric acid), the metal screen (Cr) and application of a voltage pulse (0-200V) between the liquid and the screen. (b) an SEM image of the "nanoA" nano-scale tip (scale bar = 1 µm) (c) an SEM image of the "microA" micro-scale tip (scale bar = 2 µm).

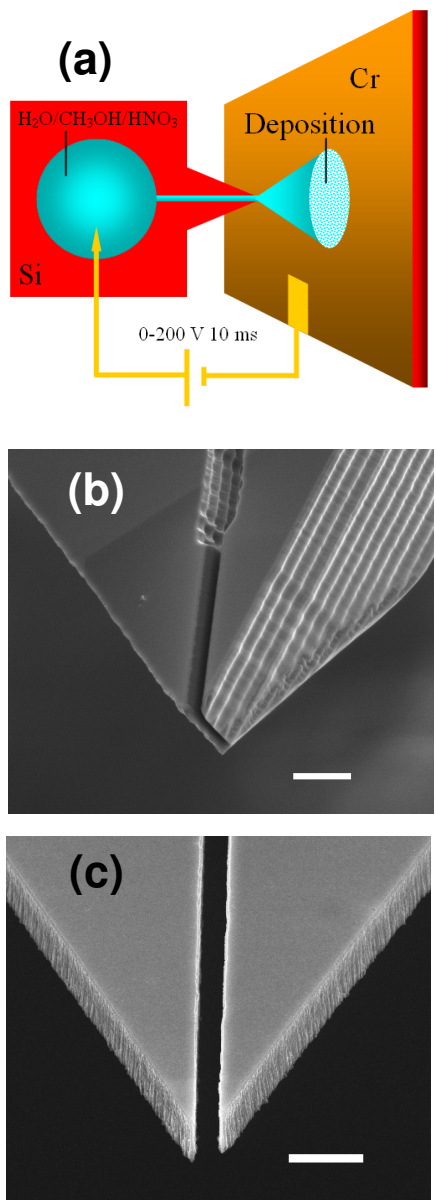

FIG. 2. Snapshot of nanodroplets. (a) revealed using AFM imaging following experiments using the "nanoA" nano-scale tip (scale bar = 5 µm). (b) zoom on a single spot (scale bar = 500 nm). (c) Plots of the spot number $n$ divided by total number of spots $N_t$ versus the spot width $w$ obtained using the nano-scale tips (red circles = nanoA; pink circles = nanoB) and two micro-scale tips 1 µm (blue circles = microA and green circles = microB) [Inset shows spot width versus spot position in trace]. (d) zoom of w versus $n/N_t$ reveals steps in the plots (black arrows). (e), histogram of spot width distribution using the nanoA tip (red bars) [Inset shows histogram for the nanoB tip (pink bars)] and (f) histogram of spot width distribution using the microA tip (blue bars) [Inset shows histogram for the microB tip (green bars)].

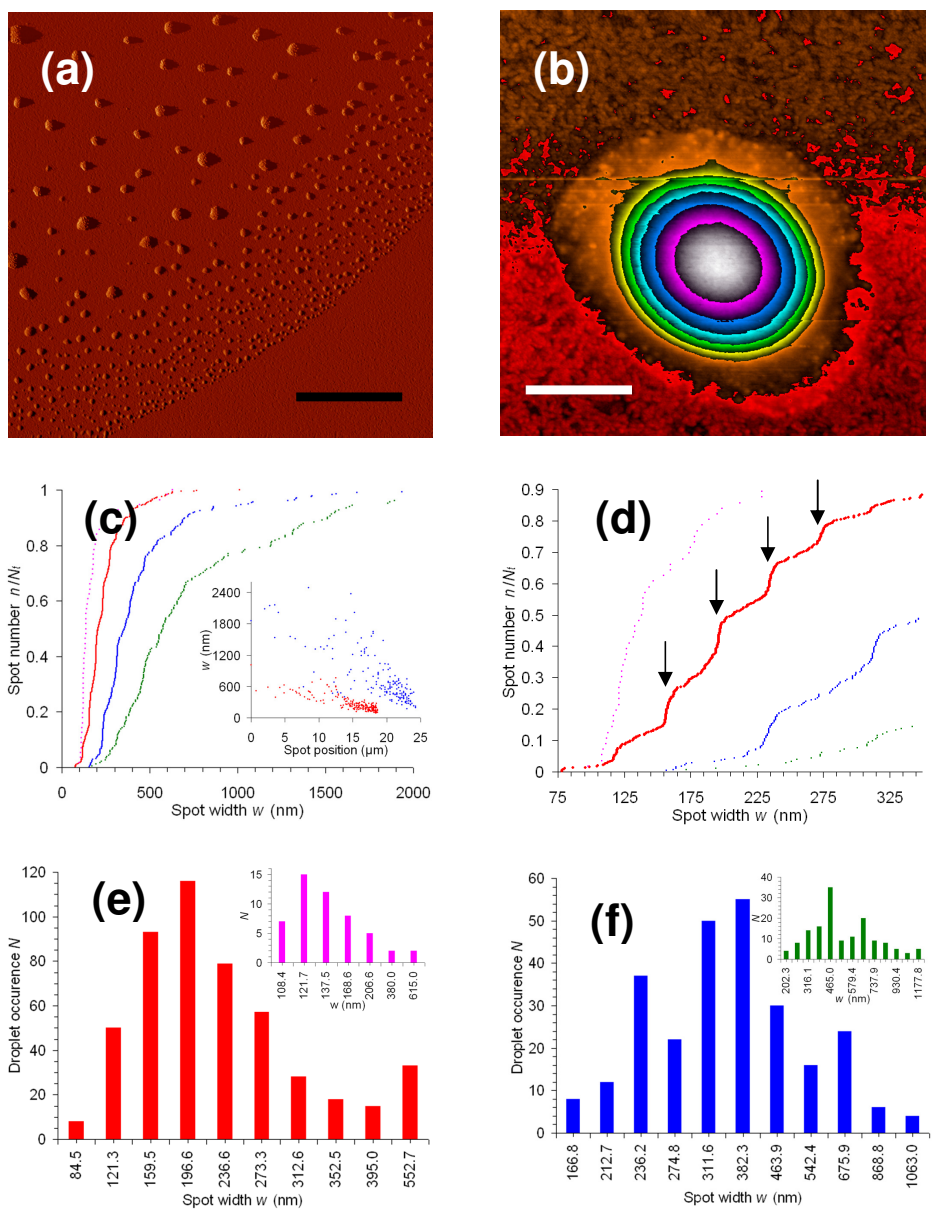

FIG. 3. Spot width $w$ and original droplet diameter $d$: schematic diagram of a droplet impinging a surface having a wetting contact angle of $\theta$.

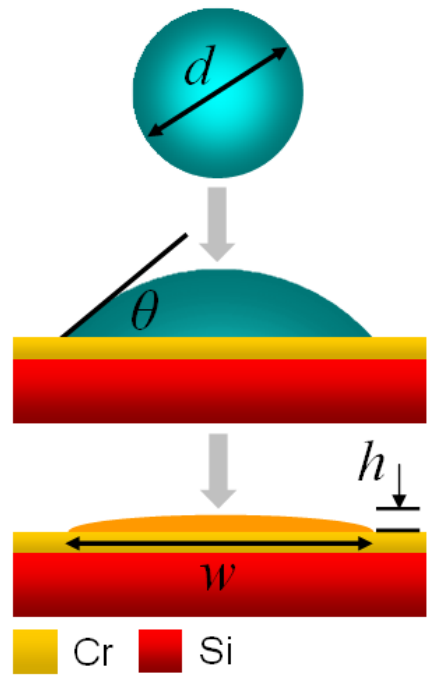

Table I. Experimental ($N_{exp}$ and $d_{exp}$) and modeling ($d_{cal}$) results for nano-scale tip "nanoA"; $d_{exp}$ corresponds to the average experimental original droplet diameter is calculated from $w_{exp}$ (see Figure 3 and text) and the droplet occurrence $N_{exp}$ is the number of spots in the given spot width interval. The black underlined bold values are the most occurring experimental. Droplet diameters corresponding to the $d_{n+1} = d_n / \sqrt[3]{2}$ model are shown in red and blue starting from the initial underlined bold values. Standard deviations are shown in brackets.

| N | $d_{exp}$ (nm) | $d_{cal}$ (nm) |
|---|---|---|
| 8 | 36.9(3.4) | 34.1 |
| 50 | 53.0(3.4) | 54.1 |
| 93 | 69.7(3.7) | 68.2 |
| 116 | **85.9(4.1)** | **85.9** |
| 79 | 103.4(3.3) | 108.2 |
| 58 | 119.4(3.5) | **119.4** |
| 28 | 136.6(3.7) | 136.4 |
| 18 | 154.0(4.1) | 150.5 |
| 15 | 172.6(2.6) | 171.8 |
| 33 | 241.5(51.3) | 238.9 |

Table II. Experimental ($N_{exp}$ and $d_{exp}$) and modeling ($d_{cal}$) results for nano-scale tip "nanoB" (See Table I for explanation).

| N | $d_{exp}$ (nm) | $d_{cal}$ (nm) |
|---|---|---|
| 7 | 47.4(2) | 47.7 |
| 15 | **53.2(1.5)** | **53.2** |
| 12 | 60.1(2.2) | **60.1** |
| 8 | 73.7(5.1) | 75.7 |
| 5 | 90.3(6.6) | 95.4 |
| 2 | 166.1(26.7) | 168.8 |
| 2 | 268.8(7) | 268 |

Table III. Experimental ($N_{exp}$ and $d_{exp}$) and modeling ($d_{cal}$) results for micro-scale tip "microA" (See Table I for explanation).

| $N$ | $d_{exp}$ (nm) | $d_{cal}$ (nm) |
|---|---|---|
| 8 | 72.9(2.7) | 75.6 |
| 12 | 92.9(4.5) | 95.3 |
| 37 | **103.2(2.5)** | 105.3 |
| 22 | 120.1(4.6) | **120.1** |
| 50 | 136.2(4) | 132.6 |
| 55 | **167.1(9.7)** | **167.1** |
| 30 | 202.7(7.8) | - |
| 16 | 237.0(10) | 240.2 |
| 24 | 295.4(209) | 302.6 |
| 6 | 379.7(29.2) | 381.2 |
| 4 | 464.5(13.5) | 480.3 |

Table IV Experimental ($N_{exp}$ and $d_{exp}$) and modeling ($d_{cal}$) results for micro-scale tip "microB" (see Table I for explanation).

| N | $d_{exp}$ (nm) | $d_{cal}$ (nm) |
|---|---|---|
| 8 | 72.9(2.7) | 75.6 |
| 12 | 92.9(4.5) | 95.3 |
| 37 | **103.2(2.5)** | 105.3 |
| 22 | 120.1(4.6) | **120.1** |
| 50 | 136.2(4) | 132.6 |
| 55 | **167.1(9.7)** | **167.1** |
| 30 | 202.7(7.8) | - |
| 16 | 237.0(10) | 240.2 |
| 24 | 295.4(209) | 302.6 |
| 6 | 379.7(29.2) | 381.2 |
| 4 | 464.5(13.5) | 480.3 |